\documentclass[12pt]{iopart}
\usepackage{iopams,graphicx}

\usepackage{graphicx}
\usepackage{dcolumn}
\usepackage{bm}

\begin{document} 

\title{Could the cosmic acceleration be transient?  A cosmographic evaluation}

\author{A.C.C. Guimar\~aes and J.A.S. Lima}

\address{Departamento de Astronomia,  Universidade de S\~ao Paulo,\\ 
Rua do Mat\~ao 1226, CEP 05508-090 S\~ao Paulo SP, Brazil}

\eads{\mailto{aguimaraes@astro.iag.usp.br}, \mailto{limajas@astro.iag.usp.br}}

\begin{abstract} 
A possible slowing down of the cosmic expansion is investigated through a 
cosmographic approach. 
By expanding the luminosity distance to fourth order and fitting the SN Ia data 
from the most recent compilations (Union, Constitution and Union 2), 
the marginal likelihood distributions for the deceleration parameter today 
suggest a recent reduction of the cosmic acceleration and 
indicate that there is a considerable probability for $q_0>0$.
Also in contrast to the prediction of the $\Lambda$CDM model, the cosmographic 
$q(z)$ reconstruction permits a cosmic expansion history where the cosmic 
acceleration could already have peaked and be presently slowing down, 
which would imply that the recent accelerated expansion of the Universe is a 
transient phenomenon.
It is also shown that to describe a transient acceleration the luminosity 
distance needs to be expanded at least to fourth order.
The present cosmographic results depend neither on the validity of general 
relativity nor on the matter-energy contents of the Universe.
\end{abstract}

\section{Introduction}

Independent measurements by various groups have suggested that
the recent expansion of the Universe is speeding up and not 
slowing down, as believed for many decades \cite{perlmutter_riess}. 
In other words, by virtue of some unknown mechanism, the cosmic expansion 
underwent a ``dynamic phase transition" whose main effect was to change the 
sign of the universal deceleration parameter $q(z)$ at low redshifts. 
Together with other complementary observations, such results are strongly 
indicating that we live in a flat, accelerating Universe composed of 
$\sim$ 1/3 of matter (baryonic + dark) and $\sim$ 2/3 of an exotic 
component with large negative pressure, usually named dark energy 
(see \cite{reviews} for some reviews).

However, what seems more certain is that the Universe was accelerating in 
a recent past, but that its current state of acceleration is less well 
determined \cite{shapiro}. 
Some authors, using a dynamical ansatz for the dark energy equation of state, 
have suggested that the cosmic acceleration has already peaked and that we 
are currently witnessing its slowing down \cite{2009PhRvD..80j1301S}. 
This behavior was also suggested by other works combining different methods 
and data \cite{WuYu}.

The possibility of a transient cosmic acceleration appears in several 
theoretical scenarios \cite{Others} and is also theoretically interesting 
because eternally accelerating universes (like $\Lambda$CDM) 
are endowed with a cosmological event horizon which prevents the construction 
of a conventional S-matrix describing particle interactions. 
Indeed, such a difficulty has been pointed out as a severe theoretical 
problem for any eternally accelerating Universe \cite{HKS01,CALR06}. 
Naturally, whether the acceleration of the Universe is slowing down the 
overall dynamic behavior in the future may be profoundly different from 
the one predicted by the $\Lambda$CDM evolution.

Here we investigate the question related to a possible slowing down of the 
recent accelerating stage of the Universe by using a distinct method. 
In order to access the present value of the deceleration parameter we 
consider the so-called cosmographic approach 
\cite{2004ApJ...614....1J,Visser04,Visser05,2007CQGra..24.5985C,2008PhRvD..78f3501C,2009arXiv0904.3550G,2010JCAP...03..005V,2010Ap&SS.330....7J} 
instead of assuming the validity of a specific gravitational theory together 
with a given dynamical dark energy model. 
As happens with kinematic models 
\cite{shapiro,2009arXiv0904.3550G,TurRie02,riess04,EM1,EM,rapp07,Gong,CL08,Cunha09}, the basic advantage of the cosmographic approach over dynamic models 
is that the former depends neither on the validity of general 
relativity nor on the matter-energy contents of the Universe. 
In addition, to verify the dependence of the method on a particular set of 
supernova data, we repeat our analysis by considering three different samples, 
namely, the Union \cite{Kowalski}, Constitution \cite{2009ApJ...700.1097H}, 
and Union 2 \cite{Union2} compilations.

\section{Cosmographic approach}

Following a traditional approach, the late time cosmic expansion can be 
expanded as \cite{Visser04,Weinb72}
\begin{equation}
\fl
\label{a_exp}
a(t)= 1+H_0(t-t_0) - \frac{1}{2}q_0 H_0^2(t-t_0)^2 +
\frac{1}{3!}j_0 H_0^3(t-t_0)^3 + 
\frac{1}{4!}s_0 H_0^4(t-t_0)^4 +
{\cal O}[(t-t_0)^5] ,
\end{equation}
where $H_0$, $q_0$, $j_0$ and  $s_0$ are the present day values of the Hubble, 
deceleration, jerk and snap parameters, respectively.
Similarly, the luminosity distance can also be expanded, yielding an extended 
version of the Hubble law
\begin{eqnarray}
  d_L(z) & = & \frac{c}{H_0}\left[z +
    \frac{1}{2}(1-q_0)z^2-\frac{1}{6}(1-q_0-3q_0^2+j_0)z^3 \right. \nonumber 
    \\ 
    &  & \left. + \frac{1}{24}(2-2q_0-15q_0^2-15q_0^3+10q_0j_0+5j_0+s_0)z^4
    \right] + {\cal O}(z^5),  
  \label{dL_exp}
\end{eqnarray}
where it was assumed a flat universe, as is done throughout this work.
As it appears, the truncation of the above expression is endowed with some 
convergence issues \cite{2007CQGra..24.5985C}, which is a source of concern 
since the SN Ia samples do have events at $z \gtrsim 1$.
One way to circumvent this problem of convergence at high redshift is to 
limit the supernova sample to events of lower redshift.  
Another approach is to parametrize cosmological distances by the so-called 
{\it y-redshift} \cite{2007CQGra..24.5985C}, defined by 
$y=(\lambda_0-\lambda_e)/\lambda_0$, a quantity related to the usual redshift, 
$z=(\lambda_0-\lambda_e)/\lambda_e$, by 
\begin{equation}
  y=\frac{z}{1+z}.
  \label{yz}
\end{equation}

Using the {\it y-redshift} we can expand the luminosity distance as 
\begin{eqnarray}
  d_L(y) & = & \frac{c}{H_0}\left[y +
    \frac{1}{2}(3-q_0)y^2+\frac{1}{6}(11-5q_0+3q_0^2-j_0)y^3 \right. \nonumber
    \\ 
    &  & \left. + \frac{1}{24}(50-26q_0+21q_0^2-15q_0^3+10q_0j_0-7j_0+s_0)y^4
    \right] + {\cal O}(y^5).  
  \label{dLy_exp}
\end{eqnarray}
This parametrization has the nice property that its convergence radius 
comprises the full range of past cosmological events.

In what follows, in order to fit the SN Ia data compiled by the 
Union (307 events), Constitution (397 events) and Union 2 (557 events) 
samples, we consider 
(i) the fourth order expansion of the luminosity distance in redshift, 
which we will label $d_L^{(4)}(z)$, that is, the truncation of (\ref{dL_exp}), 
and (ii) the fourth order expansion of the luminosity distance in 
{\it y-redshift}, $d_L^{(4)}(y)$, the truncation of (\ref{dLy_exp}). 
For the redshift expansion fit we use sub-samples with $z<z_{cut}$, labeling 
the expansions $d_L^{(4)}(z<z_{cut})$, which for $z_{cut}=0.5$ yields 157 events 
(51\%) for the Union, 247 events (62\%) for the Constitution, and 402 events 
(72\%) for the Union 2 compilation.

We construct the likelihood as ${\cal L}=\exp[-\chi^2/2]$, where
\begin{equation}
  \chi^2\equiv \sum_i \frac{\left[ \mu_B(i;M_B,\alpha,\beta)
      - \mu(i;H_0,q_0,j_0,s_0) \right]^2}{\sigma_\mu^2(i)},
\end{equation}
where $i$ runs through the SN Ia events, 
$\mu=5\log_{10}[d_L/(1 {\rm Mpc})]+25$ is the predicted distance modulus in 
a given model, and $\sigma_\mu$ is the total uncertainty on the 
distance moduli.
In the likelihood analysis we analytically marginalize over 
\cite{rapp07,2005PhRvD..72l3519N} the nuisance parameter containing the 
Hubble parameter ($H_0$) and the supernova absolute magnitude ($M_B$), 
and numerically maximize for two remaining nuisance parameters related 
to the supernova light curve fitting ($\alpha$ and $\beta$).
For both the Union and Constitution sets the SALT method values for 
$m_B^{max}$, $s$ (stretch) and $c$ (color) were used: 
$\mu_B=m_B^{max} - M_B + \alpha (s-1) - \beta c$.
Union 2 uses SALT2 light curve fitting: 
$\mu_B=m_B^{max} - M_B + \alpha x_1 - \beta c$, where $x_1$ is a 
free parameter that measures the deviation from the average decline rate.

\section{Current deceleration probability}

\begin{figure}
\centering
\includegraphics[scale=1.]{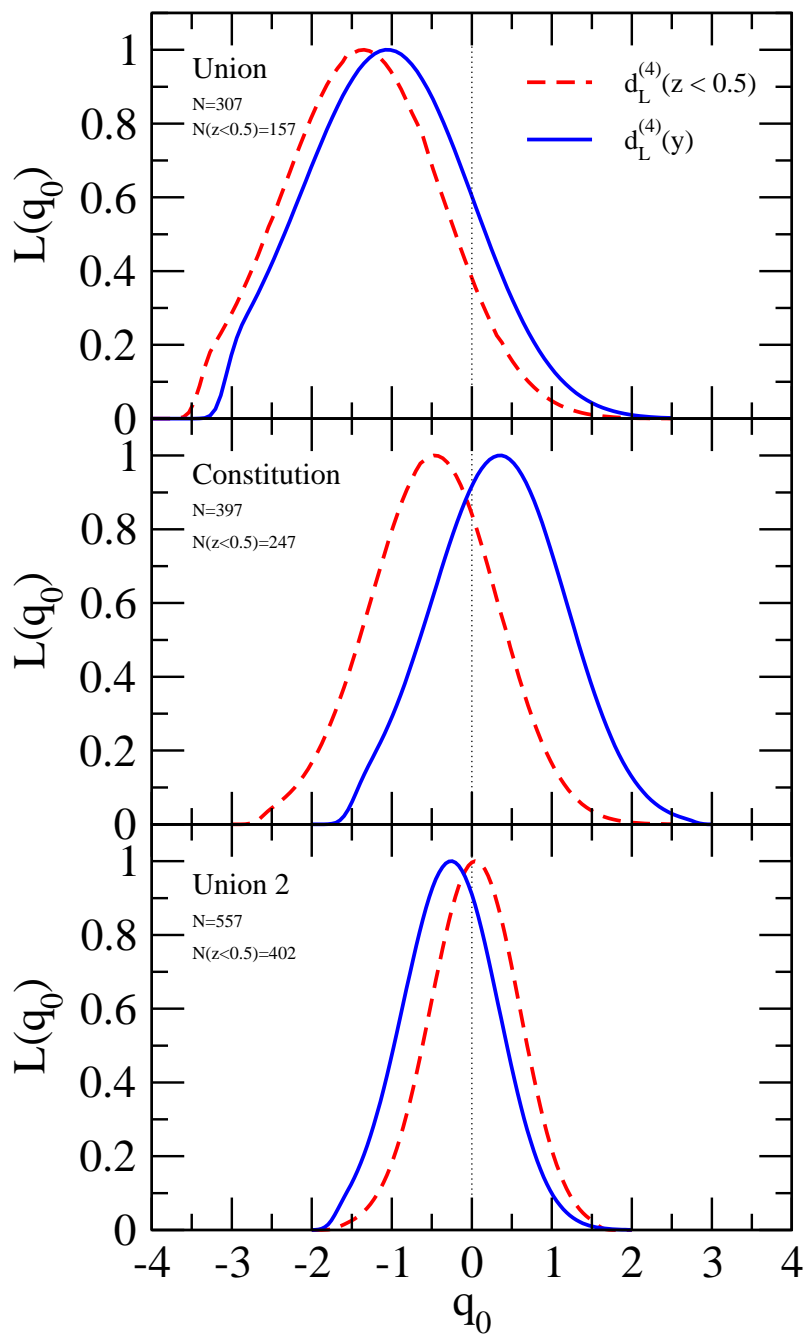}
\caption{\label{q0_likel} Marginal likelihood for the deceleration parameter 
today. Top panel uses the Union SN Ia compilation, central panel uses the 
Constitution, and the lower panel uses the Union 2. 
In the upper left corner of each panel is indicated the number of 
events in each data set.
Curves are scaled to $\max [{\cal L}(q_0)]=1$.}
\end{figure}

Since we are mainly interested in the value of the deceleration parameter 
today, we show in figure \ref{q0_likel} the marginal likelihoods for $q_0$ 
(marginalization over $j_0$ and $s_0$).
Both cosmographic parametrizations, $d_L^{(4)}(z<0.5)$ 
\footnote{The choice of $z_{cut}=0.5$ here is meant to be conservative in terms of possible convergence problems in the $d_L^{(4)}(z)$ expansion.} 
and $d_L^{(4)}(y)$, are statistically compatible and have similar dispersion, 
even though the $z<0.5$ implies a considerably smaller number of events in the 
$d_L^{(4)}(z<0.5)$ analysis in relation to the full SN Ia sets used in the 
$d_L^{(4)}(y)$ analysis.
This effect was already noticed before \cite{2008PhRvD..78f3501C}, and that 
may be indicating that the {\it y-redshift} expansion is not optimal to 
maximize the Fisher information.

The basic distinction feature between the Union and Constitution sets comes 
from 90 events at low redshift ($z<0.5$). 
They have the effect of shifting the $q_0$ likelihoods to 
higher values (figure \ref{q0_likel}), thereby increasing the 
probability of a decelerating universe (table \ref{table}).
The Union 2 has an even higher percentage of low redshift events and the 
deceleration likelihood derived from $d_L^{(4)}(z<0.5)$ is further shifted to 
higher values of $q_0$.
This suggests that the low-$z$ SN Ia events are pointing to a higher $q_0$ 
than the full sample. 
Such a trend could help to explain why Shafieloo et al. 
\cite{2009PhRvD..80j1301S} found a slowing down of the cosmic acceleration 
when analyzing the Constitution data set, but probably would find (i) a 
weaker result had they used the first Union compilation and (ii) maybe an 
even stronger result had they used the Union 2.

\begin{table}
\caption{Probability of a positive deceleration today, $P(q_0>0)$.} 
\begin{center}
\begin{tabular}{llll}
\hline 
fit $\backslash$ set & Union & Constitution & Union 2 \\
$d_L^{(4)}(z)$ \footnotemark  & 0.006 & 0.097        & 0.008   \\
$d_L^{(4)}(z<0.9)$    & -     & -            & 0.082   \\
$d_L^{(4)}(z<0.7)$    & -     & -            & 0.40    \\
$d_L^{(4)}(z<0.5)$    & 0.08  & 0.27         & 0.55    \\
$d_L^{(4)}(z<0.3)$    & -     & -            & 0.57    \\
$d_L^{(4)}(y)$        & 0.15  & 0.65         & 0.33    \\
\hline 
\end{tabular}
\end{center}
\label{table}
\end{table}
\footnotetext{The use of $z>1$ SNe in the $d_L^{(4)}(z)$ parametrization incurs more severe convergence problems, therefore these results must be taken with greater caution.}

\begin{figure}
\centering
\includegraphics[scale=0.9]{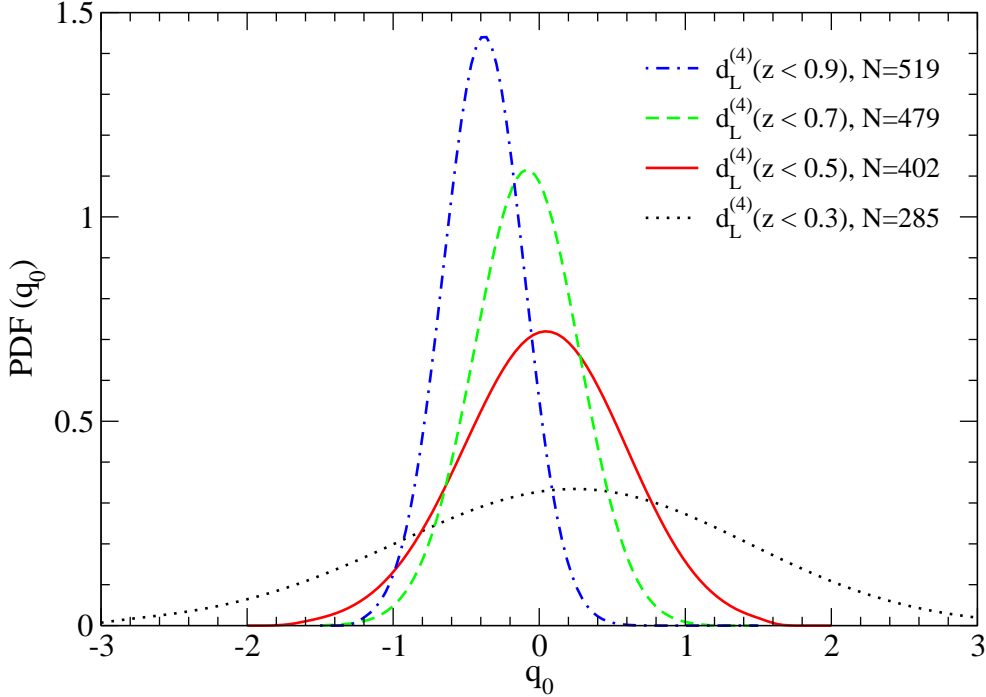}
\caption{\label{q0_zcut} Probability distribution function for the 
deceleration parameter today for the $d_L^{(4)}(z<z_{cut})$ parametrization. 
The internal legend shows the cut redshift and the corresponding number of 
SN Ia from the Union 2 compilation.}
\end{figure}

To further explore the suggestion that low-$z$ SN Ia events may be pointing 
to higher $q_0$ values, we examine the Union 2 data set up to various 
redshift cuts, $z_{cut}$, with the $d_L^{(4)}(z<z_{cut})$ parametrization.
This point of view is reinforced by the results shown in 
figure \ref{q0_zcut} and table \ref{table}.  
The more recent the SN Ia events are, the higher is the peak and average 
deceleration values today and the larger is the probability for positive 
$q_0$.
Note that the higher deceleration probability with lower redshift cutoff 
is due not only to a higher peak and average deceleration parameter, but also 
to a higher variance in the distribution, consequence of looser 
constraining of the parameter space. 
The use of various redshift cuts in the $d_L^{(4)}(y)$ parametrization also 
corroborates those findings (numerical results not shown).

In table \ref{table}, by using $d_L^{(4)}(z<z_{cut})$ and $d_L^{(4)}(y)$, 
we quantify the probability for the universe to be slowing down today for 
the three SN Ia samples.  
It has been estimated by computing the integrated marginal likelihood for 
$q_0>0$ 
\begin{equation}
  P(q_0>0)=\frac {\int_0^\infty{\cal L}(q_0) dq_0}
  {\int_{-\infty}^\infty{\cal L}(q_0) dq_0}.
\end{equation}
When just $z<0.5$ events are fitted, then the Constitution and 
Union 2 sets indicate a considerable probability of a slowing down universe. 
This probability is also large (and even dominant in the Constitution case) 
when the samples are fitted by $d_L^{(4)}(y)$.
Using a fifth order expansion, 
John \cite{2004ApJ...614....1J,2010Ap&SS.330....7J} 
also found a significant probability of a positive $q_0$.

It is not totally surprising that both parametrizations, $d_L^{(4)}(z<z_{cut})$ 
and $d_L^{(4)}(y)$, lead to distributions for $q_0$ that are not identical. 
Previous works (e.g. \cite{2009arXiv0904.3550G}) showed that best-fit 
values are model (parametrization) dependent, even for models with same 
goodness of fit.

Because the $d_L^{(4)}(z<0.5)$ and $d_L^{(4)}(y)$ fits yield broad marginal 
likelihoods that are nearly symmetrical and peaked around $q_0=0$, 
most notably with the Constitution and Union 2 compilations, 
$P(q_0>0)$ is very sensitive to the point of maximum likelihood, which is not 
a robust feature of those fits in relation to the various data sets.
This explains the variability of the results shown in table \ref{table} in 
relation to the different SN Ia sets.

\begin{figure}
\centering
\includegraphics[scale=0.6]{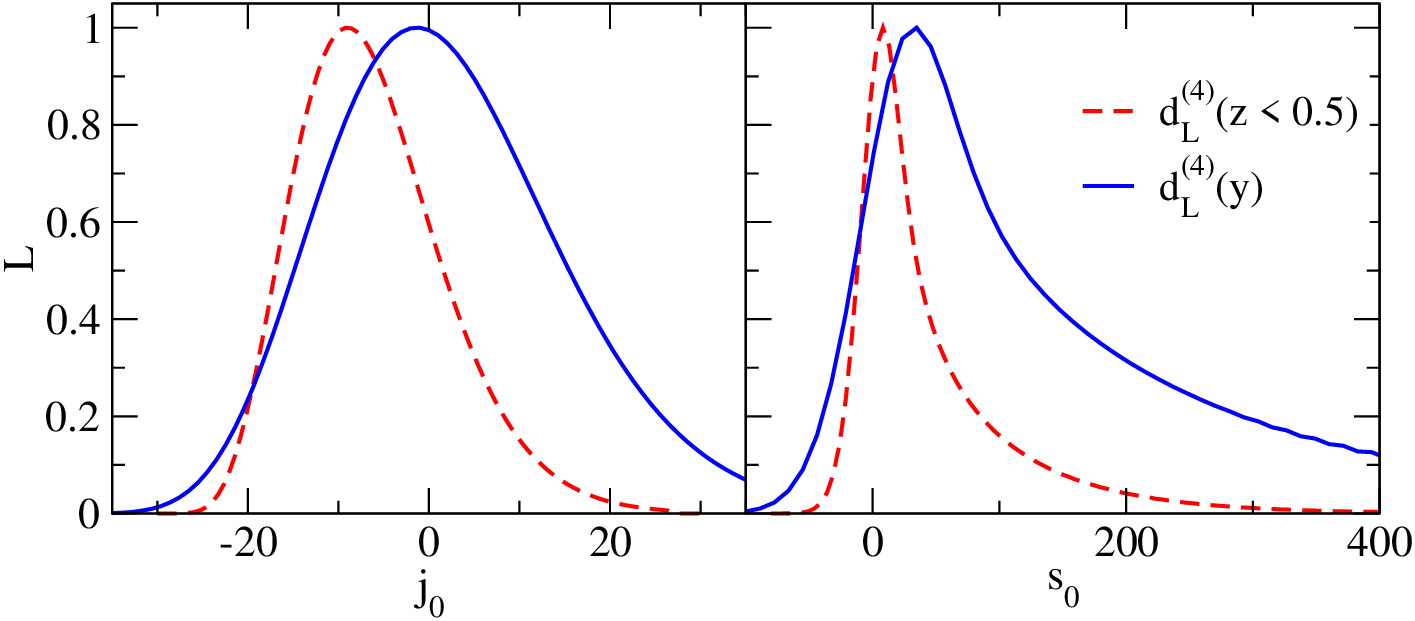}
\caption{\label{j0_s0__likel} Marginal likelihood for the jerk and snap 
parameters today from the Union 2 data.}
\end{figure}

As a byproduct, we also obtain the marginal likelihood distributions for the 
jerk and snap parameters today (figure \ref{j0_s0__likel}) from the Union 2 
data.
The $d_L^{(4)}(z<0.5)$ parametrization yields somewhat tighter constrains than 
$d_L^{(4)}(y)$, even though it uses less events.
Our results for the {\it y-redshift} parametrization are also in general 
agreement with what \cite{2010JCAP...03..005V} found using the first Union 
data set.
For reference, the flat $\Lambda$CDM model has 
$q_0=\frac{3}{2}\Omega_m-1$, 
$j_0=1$ and 
$s_0=1-\frac{9}{2} \Omega_m$. 
Therefore this model has a single free parameter, which the best fit to the 
Union 2 data gives $q_0=-0.565 \pm 0.032$. 

\section{Cosmographic deceleration history reconstruction}

Independently of the cosmic deceleration today, it is of interest to know 
if there is another change in the sign of the cosmic acceleration after a 
prior transition from a decelerated to an accelerated phase at moderate 
redshifts ($z_t\sim0.5-1$). 
From the power series expansion of the scale factor (\ref{a_exp}) one can 
also express the deceleration parameter as a power series in time. 
This time variable can be written as a power series in redshift or 
{\it y-redshift}, yielding respectively
\begin{equation}
\fl
  q(z)= q_0 + (-q_0-2q_0^2+j_0)z + 
  \frac{1}{2}(2q_0+8q_0^2+8q_0^3-7q_0j_0-4j_0-s_0)z^2 + {\cal O}(z^3) ,
  \label{q_quadr}
\end{equation}
and
\begin{equation}
\fl
  q(y)= q_0 + (-q_0-2q_0^2+j_0)y + 
  \frac{1}{2}(4q_0+8q_0^3-7q_0j_0-2j_0-s_0)y^2 + {\cal O}(y^3) .
  \label{qy_quadr}
\end{equation}
Equivalently, if the truncation of (\ref{a_exp}) is made at third order, 
then the resulting expression for $q(z)$ and $q(y)$ is linear.
Several works \cite{riess04,2009arXiv0904.3550G,EM,rapp07,Cunha09} were 
based on the $q(z)=q_0+q_1 z$ parametrization for the deceleration, 
which does not allow a transient cosmic acceleration. 
The derivation of a power expansion for the deceleration from 
the scale factor expansion establishes a natural link with the kinematic 
studies based on a $q(z)$ expansion in power series.
J. V. Cunha (private communication) employed the
$q(z)=q_0+q_1 z+ \frac{1}{2} q_2 z^2$ parametrization and found a 
suggestion of transient acceleration (see \cite{Gong} for another 
formulation of a quadratic kinematic model).
The quadratic form of (\ref{q_quadr}) and (\ref{qy_quadr}) 
allows for a decelerated past, a transition to an accelerated phase, 
a point of maximum acceleration, then a slowing down of the acceleration and 
a transition to a recent or future decelerating phase. 
Similar behavior of transient acceleration is predicted or allowed by several 
dynamic models \cite{2009PhRvD..80j1301S,CALR06}.
In contrast, the $\Lambda$CDM model predicts a monotonic deceleration history 
connecting its asymptotic limits in the past and future, 
$q(z\rightarrow \infty)=0.5$ and $q(z\rightarrow -1)=-1$.

\begin{figure}
\centering
\includegraphics[scale=1.]{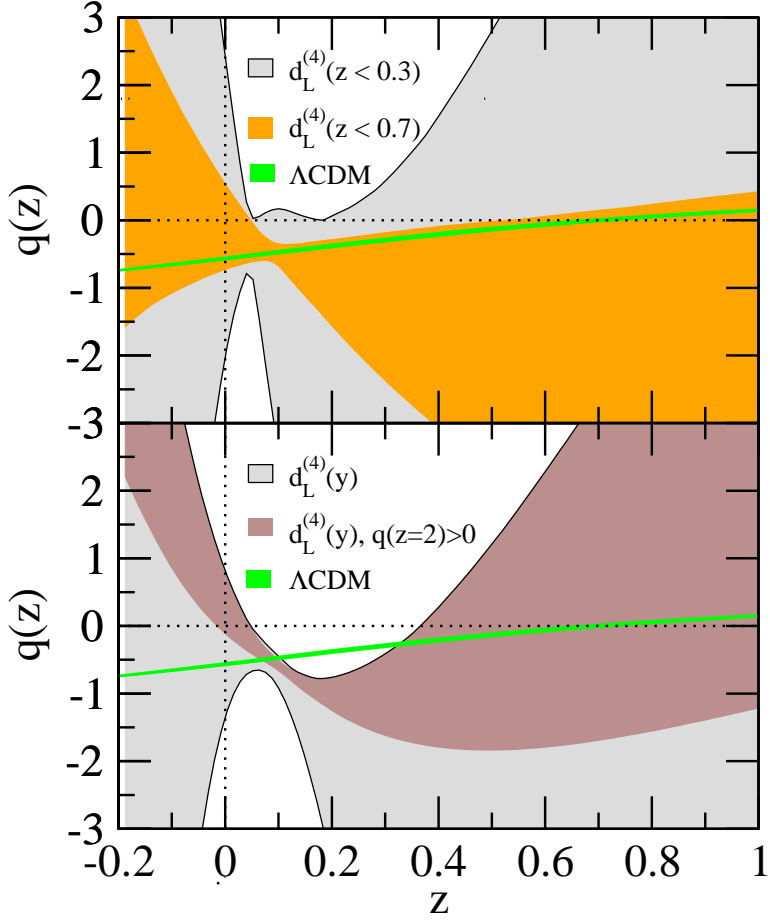}
\caption{\label{qXz} Deceleration history reconstruction. 
The gray area and the darker area inside it delimit the 1$\sigma$ 
confidence region for the $q(z)$ reconstruction obtained from various 
Union 2 SN Ia analyzes.
Top panel: $d_L^{(4)}(z<0.3)$ (gray),  
$d_L^{(4)}(z<0.7)$ (darker internal area) and flat 
$\Lambda$CDM model (green thinner area).
Bottom panel: $d_L^{(4)}(y)$ with no extra conditions (gray) and with the 
condition that the deceleration parameter must be positive at $z=2$ inside 
the 1$\sigma$ confidence region (darker internal area).}
\end{figure}

Even though (\ref{q_quadr}) and (\ref{qy_quadr}) are expansions of same 
order in their respective variables, the second is a more precise expression 
since the order of its error ${\cal O}(y^3)={\cal O}(z^3/[1+z]^3)$ is always 
lower than the order of the error for the first, ${\cal O}(z^3)$, in the past.

We use the cosmographic parameters ($q_0$, $j_0$ and $s_0$) obtained in 
each parametrization with the Union 2 sample to reconstruct $q(z)$ using 
(\ref{qy_quadr}) and (\ref{yz}), which is shown in figure \ref{qXz}.
The confidence regions shown in the figure are drawn from all $q(z)$ curves 
allowed by the parameter values such that 
${\cal L}(q_0, j_0 ,s_0 )/{\cal L}_{max}$ is less than a threshold
\cite{2010LNP...800..147V}.
The top panel shows the deceleration reconstruction 
for the $d_L^{(4)}(z<z_{cut})$ parametrization for $z_{cut}=0.3$ and $z_{cut}=0.7$.
The first constrains very loosely $q(z)$, not even showing a phase of 
accelerated cosmic expansion with confidence above 1$\sigma$.
The higher cut redshift allows a tighter constraining, being maximum for 
$z\sim 0.1$.

The $q(z)$ reconstruction obtained from the $d_L^{(4)}(y)$ parametrization is 
shown in the bottom panel of figure \ref{qXz}. 
It defines a clear phase of accelerated expansion around $z\sim 0.2$, 
but suggests it is slowing down around $z\sim 0.1$, where the $q(z)$ 
reconstruction is maximally constrained.

For comparison, figure \ref{qXz} also plots the results for a flat 
$\Lambda$CDM model, showing that under this parametrization all 
redshifts have equally well determined deceleration.
As we know that the SN Ia data is limited to a redshift range, it is clear 
that the preponderant factor in this $\Lambda$CDM result is the model itself, 
not the data. 
We also note that, even though the comparison of the cosmographic 
parametrizations with the $\Lambda$CDM model is not the objective of this 
work, all have very similar goodness of fit as measured by the 
reduced $\chi^2$. 

All cosmographic parametrizations contain transient acceleration solutions 
in their 1$\sigma$ confidence regions with a considerable positive 
deceleration probability today and in the near future, if we assume that the 
cosmographic parametrization has some predictive power.
However, due to the large uncertainties in the $q(z)$ reconstructions at 
large redshifts or in a projection into the future ($z<0$) from the 
cosmographic parametrizations of the Union 2 data alone, it is not possible 
to affirm if there is a transition from a phase of accelerated 
expansion to a present or future phase of deceleration.

An improvement in the containment of the allowed $q(z)$ could be achieved 
by imposing external conditions to the cosmographic analysis, for example that 
in the remote past the matter dominated universe should be decelerated. 
As an exploratory investigation, we implement this external restriction to 
the analysis in the bottom panel of figure \ref{qXz} by imposing that 
$q(z=2)>0$ 
\footnote{The choice of $z=2$ is arbitrary, but meant to be a conservative 
compromise. A larger redshift would imply a lower constraint on the fitting, 
and a with a smaller redshift, the constraint would be stronger, but then we 
would be less certain about the reality of the $q(z)>0$ condition.} 
at the 1$\sigma$ confidence level. 
As a result, the allowed $q(z)$ is further constrained, only leaving transient 
acceleration solutions.

The strict transient behavior obtained with the external $q(z=2)>0$ condition 
may be a consequence of the fourth order cosmographic parametrization for the 
luminosity distance, as this implies a quadratic $q(z)$ which may be forced to 
have two zeros (two transition redshifts). 
Every parametrization implies a bias, and to test if this is the case for the 
transient solutions in the approach adopted, one would need to go to higher 
order in the cosmographic expansion.
However, models with a higher number of parameters would need more data to 
meaningfully constrain the parameter space.

\section{Conclusion}

To address the main question of this work, we should focus on the results for 
the Union 2 compilation, since it is currently the largest and most updated.
Both kinds of cosmographic parametrization, $d_L^{(4)}(z<z_{cut})$ and 
$d_L^{(4)}(y)$, allow the circumvention of convergence issues related to the 
power expansion, broadly agree and 
(i) suggest that lower redshift SN events point to lower cosmic acceleration, 
and 
(ii) indicate that there is a considerable probability for the Universe to be 
decelerating today.
The slowing down of the cosmic acceleration and even a transient acceleration 
scenario is allowed in all cosmographic parametrizations. 

The possibility that those results may be due to bias induced by the 
form of the parametrizations and data analysis artifacts may be examined 
inside the cosmographic approach by going to higher order in the scale factor 
expansion and through the use of simulated data inside some conveniently chosen 
fiducial cosmological models.
That in itself would constitute an interesting follow up study that could be 
useful to improve the understanding of the limitations of the observational 
data sets and of the analysis methods. 

What may be causing a possible slowing down of cosmic acceleration and 
transient cannot be answered with basis only in a cosmographic approach.  
We recall that in the accepted scenario (inflation followed by radiation, 
dark matter and dark energy dominated stages), the Universe already had two 
accelerating stages and what is being discussed is whether the present 
supernova data are compatible with a new transition to a decelerating phase. 
Some examples of dynamical scenarios discussed in the literature include 
quintessence models in which the equation of state parameter becomes positive 
in the future, and some D-branes inspired scenarios, among others 
\cite{Others,CALR06}. 
Naturally, we cannot exclude the possibility that such a slowing down might 
be provoked by some exotic kinematic (or dynamic effect) in the presence of 
inhomogeneities \cite{Buchert}.

In conclusion, we used fourth order expansions of the luminosity distance to 
describe the recent cosmic expansion as probed by SN Ia observations. 
We showed that one needs to go at least to fourth order in the expansion of 
the luminosity distance to be able to describe a transient acceleration. 
Other works \cite{2008PhRvD..78f3501C,2010JCAP...03..005V}, using smaller 
data sets than some of those used here, indicate that going to higher orders 
does not add meaningful statistical information.
Under a cosmographic analysis of the most recent SN Ia compilations, we have 
found a suggestion of cosmic acceleration slowing down and the existence of 
a considerable probability in the relevant parameter space that the Universe 
is already in a decelerating expansion regime.
This result is in great contrast to the standard $\Lambda$CDM model, 
which predicts that the cosmic expansion is accelerating and must remain 
so forever with an ever increasing acceleration.

The predictability of the cosmographic parametrizations are considerably 
inferior to the $\Lambda$CDM model and other dynamical models, but that 
seems to be the price for avoiding the usual assumptions about the physical 
contents of the Universe and the underlying gravity theory.
It should also be stressed that the price for avoiding convergence issues 
for the expansion in powers of $z$ has severe consequences on model 
predictability, but this could be overcome by a much larger SN Ia sample 
(which is foreseen in future surveys), by the use of other kinds of data and 
priors, and maybe by a smarter (optimal) cosmographic modeling.

\section*{Acknowledgments}
The authors thank V. Vitagliano and the anonymous referees for their 
useful comments.
ACCG is supported by FAPESP under grant 07/54915-9. JASL is partially 
supported by CNPQ (grant 304792/2003-9) and FAPESP (grant 04/13668-0).



\section*{References}
\begin {thebibliography}{X}
\bibitem{perlmutter_riess} 
A. G. Riess {\it et al.}, Astron. J. {\bf 116}, 1009 (1998);
S. Perlmutter {\it et al.}, Nature {\bf 391}, 51 (1998); 
S. Perlmutter {\it et al.}, Astrophys. J. {\bf 517}, 565 (1999).
\bibitem{reviews} P.~J.~E.~ Peebles  and B. Ratra, Rev.~Mod.~Phys. {\bf 75} 559 (2003); 
T. Padmanabhan, Phys.~Rept. {\bf 380}, 235 (2003); 
J.~A.~S. Lima,  Braz.~Journ.~Phys. {\bf 34}, 194 (2004), [astro-ph/0402109]; 
E.~J. M. Copeland, S. Tsujikawa,  Int.~J.~Mod.~Phys. {\bf D15}, 1753 (2006); 
J.~A. Frieman,  M. S.  Turner, and D. Huterer, Ann.~Rev.~Astron. \& Astrophys. {\bf 46}, 385 (2008).
\bibitem{shapiro} C. Shapiro, M. S. Turner, Astrophys. J. {\bf 649}, 563 (2006).
\bibitem{2009PhRvD..80j1301S} A. Shafieloo, V. Sahni, A. A. Starobinsky, Phys. Rev. D {\bf 80}, 101301 (2009). 
\bibitem{WuYu} 
Z. Li, P. Wu, H. Yu, JCAP, 11, 31 (2010);
Z. Li, P. Wu, H. Yu, Physics Letters B, 695, 1 (2011);
P. Wu, H. Yu, arXiv:1012.3032 (2010)
\bibitem{Others} 
J. D. Barrow, R. Bean, J. Magueijo, MNRAS, 316, L41 (2000);
M. C. Bento, O. Bertolami, N. C. Santos, Phys. Rev. D {\bf 65}, 067301 (2002);
R. Kallosh, A. Linde, S. Prokushkin, M. Shmakova, Phys. Rev. D {\bf 66}, 123503 (2002);
R. Kallosh, A. Linde, 2003, JCAP {\bf 2}, 2 (2003);
U. Alam, V. Sahni, A. A. Starobinsky, JCAP {\bf 4}, 2 (2003);
V. Sahni, Y. Shtanov, JCAP {\bf 11}, 14 (2003);
D. Blais, D. Polarski, Phys. Rev. D {\bf 70}, 084008 (2004);
S. K. Srivastava, Physics Letters B {\bf 648}, 119 (2007).
J. S. Alcaniz, H. {\v S}tefan{\v c}i{\'c}, Astropart. Phys. {\bf 462}, 443 (2007); 
M. C. Bento, R. G. Felipe, N. M. Santos, Phys. Rev. D {\bf 77}, 123512 (2008);
J. C. Fabris, B. Fraga, N. Pinto-Neto and W. Zimdahl, JCAP {\bf 4}, 8 (2010).
\bibitem{HKS01} S. Hellerman, N. Kaloper and L. Susskind, J. High Energy Phys. {\bf 06}, 003 (2001); J. M. Cline, J. High Energy Phys. {\bf 08}, 35  (2001).
\bibitem{CALR06} F. C. Carvalho, J. S. Alcaniz, J. A. S. Lima, R. Silva, Physical Review Letters {\bf  97}, 081301 (2006); J. S. Alcaniz, arXiv:0911.1087 (2009).
\bibitem{2004ApJ...614....1J} M. V. John, Astrophys. J. {\bf 614}, 1 (2004).
\bibitem{Visser04} M. Visser, Class. Quant. Grav. {\bf 21}, 2603 (2004).
\bibitem{Visser05} M. Visser, General Relativity and Gravitation {\bf 37} 1541 (2005).
\bibitem{2007CQGra..24.5985C} C. Catto{\"e}n, M. Visser, Classical and Quantum Gravity {\bf 24}, 5985 (2007).
\bibitem{2008PhRvD..78f3501C} C. Catto{\"e}n, M. Visser, Phys. Rev. D {\bf 78}, 063501 (2008).
\bibitem{2009arXiv0904.3550G} A. C. C. Guimar{\~a}es, J. V. Cunha, J. A. S. Lima, JCAP {\bf 10}, 010 (2009).
\bibitem{2010JCAP...03..005V} V. Vitagliano, J.-Q. Xia, S. Liberati, M. Viel, JCAP {\bf 3}, 5 (2010).
\bibitem{2010Ap&SS.330....7J} M. V. John, Astrophys. Space Sci., 330, 7 (2010)
\bibitem{TurRie02} M. S. Turner, A. G. Riess, Astrophys. J. {\bf{569}}, 18 (2002).
\bibitem{riess04} A. G. Riess {\it et al.}, Astrophys.\ J.\  {\bf 607}, 665 (2004).
\bibitem{EM1} {\O}. Elgar{\o}y, T. Multam\"{a}ki, Mon.\ Not.\ Roy.\ Astron.\ Soc.\  {\bf 356}, 475 (2005).
\bibitem{EM} {\O}. Elgar{\o}y, T. Multam{\"a}ki, JCAP {\bf 9} 2 (2006).
\bibitem{rapp07} D. Rapetti, S. W. Allen, M. A. Amin, R. D. Blandford, MNRAS {\bf 375} 1510 (2007).
\bibitem{Gong} Y. Gong and A. Wang, Phys. Rev. D {\bf{75}}, 043520 (2007).
\bibitem{CL08} J. V. Cunha, J. A. S. Lima, Mon. Not. R. Astron. Soc. {\bf{390}} 210 (2008).
\bibitem{Cunha09} J. V. Cunha, Phys. Rev. D {\bf 79} 047301 (2009).
\bibitem{Kowalski} M. Kowalski {\it et al.},  Astrophys. J. {\bf 686} 749 (2008).
\bibitem{2009ApJ...700.1097H} M. Hicken, W. M. Wood-Vasey, S. Blondin, P. Challis, S. Jha, P. L. Kelly, A. Rest, R. P. Kirshner, Astrophys. J. {\bf 700}, 1097 (2009).
\bibitem{Union2} R. Amanullah {\it et al.}, Astrophys. J. {\bf 716}, 712 (2010).
\bibitem{Weinb72} S. Weinberg, {\it Cosmology and Gravitation}, John Wiley Sons, New York (1972).
\bibitem{2005PhRvD..72l3519N} S. Nesseris, L. Perivolaropoulos, Phys. Rev. D {\bf 72}, 123519 (2005).
\bibitem{2010LNP...800..147V} Verde, L.\ 2010, Lecture Notes in Physics, Berlin Springer Verlag, 800, 147. 
\bibitem{Buchert} T. Buchert, Gen. Rel. Grav. {\bf 40}, 467 (2008); C. G. Tsagas, MNRAS  {\bf 405}, 503 (2010).
\end{thebibliography}
\end{document}